\documentclass[preprintnumbers,amsmath,amssymb,pre]{revtex4}
\usepackage{citesort}
\usepackage{amsmath,amssymb}
\usepackage{epsfig}
\usepackage{graphicx}
\usepackage{array}
\usepackage{subfig}

\begin{document}

\title{Universal functions and exactly solvable chaotic systems}
\author{M\'onica A. Garc\'ia-\~Nustes\footnote{Corresponding author. Fax: +58-212-5041566; e-mail: mogarcia@ivic.ve}}
\address{Centro de F\'{\i}sica, Instituto Venezolano de Investigaciones Cient\'ificas, Apartado 21827, Caracas 1020-A, Venezuela}
\author{Emilio Hern\'andez-Garc\'{\i}a}
\address {Departamento de F\'{\i}sica Interdisciplinar,Instituto Mediterráneo de Estudios Avanzados CSIC-Universidad de las Islas Baleares, E-07122-Palma de Mallorca, Espa\~na}
\author{Jorge A. Gonz\'alez}
\address{Centro de F\'{\i}sica, Instituto Venezolano de Investigaciones Cient\'ificas, Apartado 21827, Caracas 1020-A, Venezuela}

\begin{abstract}
A universal differential equation is a nontrivial differential equation the solutions of which approximate to arbitrary accuracy any continuous function on any interval of the real line. On the other hand, there has been  much interest in exactly solvable chaotic maps. An important problem is to generalize these results to continuous systems.
 Theoretical analysis would allow us to prove theorems about these systems and predict new phenomena. In the present paper we discuss the concept of universal functions and their relevance to the theory of universal differential equations. We present a connection between universal functions and solutions to chaotic systems. We will show the statistical independence between $X(t)$ and $X(t + \tau)$ (when $\tau$ is not equal to zero) and $X(t)$ is a solution to some chaotic systems. We will construct universal functions that behave as delta-correlated noise. We will construct universal dynamical systems with truly noisy solutions. We will discuss physically realizable dynamical systems with universal-like properties.
\end{abstract}
\maketitle

\section{Introduction}
Recently there has been great interest in exactly solvable chaotic systems \cite{29,30,31,32,33,34,35}.
S. Ulam and J. von Neumann were the first to prove that the general solution to the logistic map can be found \cite{29,30}.
It is very important to extend these results to continuous systems.
Theoretical analysis would allow us to prove theorems about these systems and predict new phenomena.
Another surprising parallel development is that of ``universal differential equations" .
A universal differential equation is a nontrivial differential-algebraic equation with the property that its solutions approximate to arbitrary accuracy any continuous function on any interval of the real line \cite{Rubel1,BuckR,Duffin,Rubel2,Bosher1,Bosher2,Elsner,Rubel3,Rubel4}.
Rubel found the first known universal differential equation by showing that there are differential equations of low order (e.g. fourth-order) which have solutions arbitrarily close to any prescribed function \cite{Rubel1}. The existence of universal differential equations illustrate the amazing complexity that solutions of low-order dynamical systems can have.
In the present paper, we will review some developments in the areas of universal differential equations and exactly solvable chaotic dynamical systems.
We will discuss the concept of ``universal functions" and their relevance to the theory of universal differential equations.
We will show the statistical independence between $X(t)$ and $X(t+\tau)$ (when $\tau\neq0$) and $X(t)$ is the solution to some chaotic dynamical systems.
We will present a connection between universal functions and solutions to chaotic systems.
We will construct universal functions that behave as $\delta$-correlated noise. We will construct universal differential equations with truly noisy solutions.
We will discuss physically realizable dynamical systems with universal properties and their potential applications in secure communications and analog computing.

\section {Universal differential equations}

Rubel's theorem \cite{Rubel1} is:

There exists a nontrivial fourth-\,order differential equation
\begin{equation}
\label{EqGeneral}
P (y{'}, y{''}, y{'''}, y{''''}) = 0,
\end{equation}
where $y^{'}= \frac{dy}{dt}$, $P$ is polynomial in four variables, with integer coefficients, such that for any continuous function $\phi(t)$ on $(-\infty,\infty)$ and for any positive continuous function $\varepsilon (t)$ on $(-\infty, \infty)$, there exists a $C^{\infty}$ solution $y(t)$ such that
$\vert y(t) - \phi(t)\vert \ < \varepsilon (t)$ for all $t$ on $(-\infty, \infty)$.
A particular example of Eq.(\ref{EqGeneral}) is the following:
\begin{equation}
\label{EqExample}
\begin{split}
&3\,y{'}^{4} y{''} y{''''}^{2} - 4\,y{'}^{4}y{'''}^{2} y{''''} + 6\, y{'}^{3}y{''}^{2} y{'''}
y{''''} + 24 y{'}^{2} y{''}^{4} y{''''}\\ & -
29\,y{'}^{2}y{''}^{2} y{'''}^{2} - 12\, y{'}^{3} y{''} y{'''}^{3} + 12\, y{''}^{7} = 0,
\end{split}
\end{equation}

Duffin \cite{Duffin} has found two additional families of universal differential equations:
\begin{equation}
\label{EqDuff1}
m  y{'}^{2}y{''''} + (2-3 m)y{'}y{''}y{'''} + 2 (m-1) y{''}^{3} = 0
\end{equation}
and
\begin{equation}
\label{EqDuff2}
m^{2}y{'}^{2}y{''''} + 3m (1-m)   y{'}y{''}y{'''} + (2m^{2} - 3 m+1) y{''}^{3} = 0
\end{equation}
 where $m > 3$.

Recently, Briggs \cite{Briggs} has found a new family:
\begin{equation}
\label{EqBriggs1}
y{'}^{2}y{''''}  - 3  y{'}y{''}y{'''} + 2 (1- n^{-2})y{''}^{3} = 0,
\end{equation}
for $n > 3$.

The solutions are $C^{n}$.

We would like to make some observations about these equations.

Rubel's function $y(t)$ in Eq.(\ref{EqExample}) is $C^{\infty}$ but not real-analytic, typically having a countable number of essential singularities.

The functions used to reconstruct the solutions to Eq.(\ref{EqExample}) are of the form
\begin{equation}
y = A f(\alpha t + \beta) + B,
\end{equation}
where $f(t) = \int g(t) dt$, $g(t) = \exp\left[-\frac{1}{(1 - t^{2})}\right]$ for  $-1 < t < 1$, with $g(t) = 0$ for all other $t$.

The solutions to equations (\ref{EqDuff1}) are trigonometric splines. The kernel $g(t)$ is defined as
\begin{equation}
g(t) = a[\cos bt + c]^{m}.
\end{equation}

On the other hand, the kernel employed to obtain the solutions to Eq. (\ref{EqDuff2}) are polynomial splines:
\begin{equation}
g (t) = a [1 - (bt + c)^{2}] ^{m}.
\end{equation}
\begin{figure}
\centering
\subfloat[]{
\label{fig:DiffEqDuffin1:a}
\includegraphics[width=0.28\linewidth]{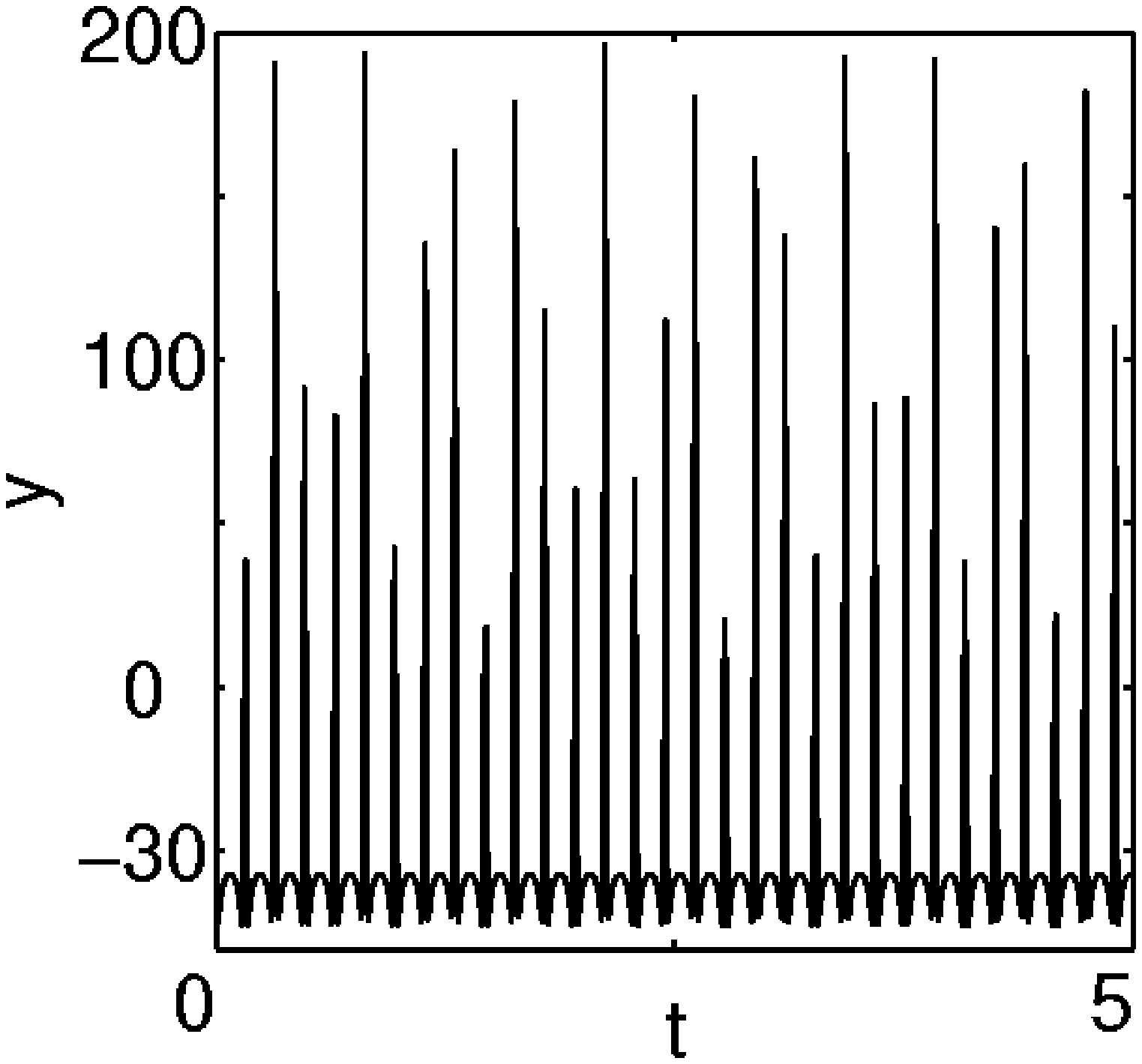}}
\quad
\subfloat[]{
\label{fig:DiffEqDuffin2:b}
\includegraphics[width=0.27\linewidth]{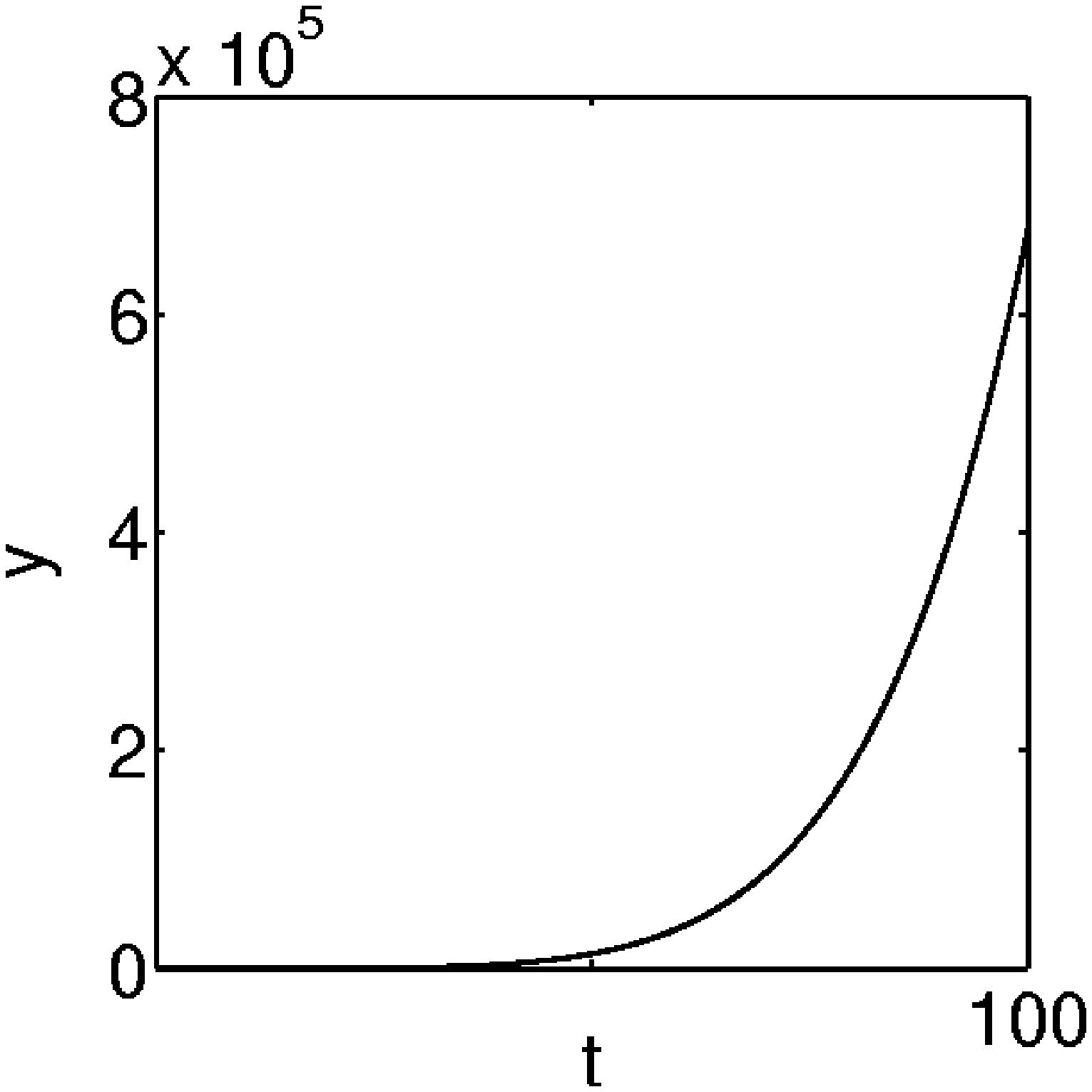}}
\caption{ Solution $y(t)$ for \;(a)\;equation (\ref{EqDuff1}), and \;(b)\;equation (\ref{EqDuff2}) with $m=4$. }
\label{fig:DiffEqDuffint}
\end{figure}

The solutions are very unstable. This phenomenon can be observed in Fig.(\ref{fig:DiffEqDuffin1:a}) and Fig.(\ref{fig:DiffEqDuffin2:b}).

\section{Exactly solvable chaotic maps}

Ulam and von Neumann \cite{29,30} proved that the function $X_{n} = \sin^{2}(\theta \pi\,2^{n})$ is the general solution to the logistic map
\begin{equation}
X_{n+1} = 4X_{n} (1 - X_{n}).
\end{equation}

Recently, many papers have been dedicated to exactly solvable chaotic maps \cite{31,32,33,34,35,BeckC,GroupNazar,GroupChaos,GroupPhysLettA}. In some of these papers, the authors not only find the explicit functions $X_{n}$ that solve the maps, but also they discuss the statistical properties of the sequences generated by the chaotic maps.
For many of the maps discussed in the mentioned papers, the exact solution can be written as
\begin{equation}
X_{n} = P (\theta k^{n}),
\end{equation}
where $P(t)$ is a periodic function, $\theta$ is a fixed real parameter and $k$ is an integer.

For instance, $X_{n} = \cos (2\pi\theta\,3^{n})$ is the solution to map $X_{n+1} = X_{n}(4X_{n}{}^{2} - 3)$. This is a particular case of the Chebyshev maps \cite{33}.

Very interesting chaotic maps of type $X_{n+1} = f(X_{n})$ can be constructed when function $P(t)$ is a combination of trigonometric, elliptic and other generalized periodic functions \cite{34}.

\section{Statistical independence}
\label{Sec:StatIndep}

We will consider statistical independence in the sense of M. Kac \cite{Kac1,Kac2,Kac3,Kac4}. In this framework, two functions $f_{1}(t), \; f_{2}(t)$ are independent if the proportion of time during which, simultaneously, $f_{1}(t)< \alpha_{1}$ and $f_{2} (t) < \alpha_{2}$ is  equal to the product of the proportions of time during which separately $f_{1}(t)<\alpha_{1}, f_{2}(t) < \alpha_{2}$.

First, we will present some results about functions of natural argument $n$.

Consider a generalization to the functions that are exact solutions to the Chebyshev maps:
\begin{equation}
X_{n} = \cos(2\pi\theta Z^{n}),
\end{equation}
where $Z$ is a generic real number.

Any set of subsequences $X_{s}, X_{ s+1},\cdots, X_{s+r}$ (for any $r$) constitutes a set of statistically independent random variables.

If $E(X)$ is the expected value of quantity $X$, let us define the $r$-\,order correlations \cite{BeckC}:
\begin{equation}
E (X_{n_{1}}X_{n_{2}} \cdots X_{n_{r}}) = \int\limits^{1}_{-1}dX_{0} [\rho(X_{0}) X_{n_{1}}X_{n_{2}} \cdots X_{n_{r}}].
\end{equation}

Here $E(X_{n}) = 0$, $\;-1 \leqslant X_{n} \leqslant 1$,$\;\rho (X) = \frac{1}{\pi\sqrt{1-X^{2}}}$,
$\;X_{0} = \cos(2\pi\theta)$.

In Ref. \cite{Group1} it is shown that
\begin{equation}
\label{EqIndependence}
E (X_{s}^{n_{0}} X_{s+1}^{n_{1}}\cdots X_{s+r}^{n_{r}}) = E (X_{s}^{n_{0}}) E (X_{s+1}^{n_{1}}) \cdots E(X_{s+r}^{n_{r}})
\end{equation}
for all positive integers $n_{0}, n_{1}, n_{2}, \cdots, n_{r}$.

The results about the independence of subsequences of function $X_{n} = \cos(2 \pi \theta Z^{n})$ can be extended to more general functions as the following:
\begin{equation}
X_{n} = P(\theta \,T Z^{n}),
\end{equation}
where $P(t)$ is a periodic function, $T$ is the period of $P(t)$ and $\theta$ is a parameter \cite{Group1}.

M. Kac \cite{Kac1,Kac2,Kac3,Kac4} has studied independence of different continuous functions, e.g. $f_{1}(t) = cos (t), f_{2} (t) = cos (\sqrt{2}t)$. However, these functions are periodic \cite{Kac1,Kac2,Kac3,Kac4,Kac5}.

We are interested in the independence of continuous functions in the sense that the same function can produce statistically independent random variables if evaluated at different times. That is, the functions $f(t)$ and $f(t + \tau)$ should be statistically independent if $\tau \neq 0$.

Periodic functions will never possess this property.

Using the theorems of Refs. \cite{Kac1,Kac4}, we obtain the result that functions $f_{1} = \cos (\lambda_{1} t),f_{2} = \cos (\lambda_{2} t),\cdots, f_{r} = \cos (\lambda_{r}t)$ constitute a set of independent functions when the numbers $\lambda_{1}, \lambda_{2}, \cdots, \lambda_{r}$ are linearly independent over the rationals.
This is equivalent to the fact that if $\alpha_{1}, \alpha_{2},\cdots , \alpha_{r}$ are rational, $\alpha_{1}\lambda_{1} + \alpha_{2}\lambda_{2} + \cdots + \alpha_{r}\lambda_{r} = 0$ only if $\alpha_{1}, \alpha_{2},\cdots, \alpha_{r}$ are all zero.

Consider the following continuous function constructed as a continuous analogous to the function (11):
\begin{equation}
\label{Eq21}
X(t) = cos(\theta e^{bt}).
\end{equation}

It is not difficult to see that the functions $X(t)$ and $Y(t) = X(t + \tau)$ are independent (in M. Kac's sense) for (Lebesgue) almost all $\tau >0$. Note that $Y(s) = \cos \lambda s, X(s) = \cos s$, where $s=\theta e^{bt}$, $\lambda = e^{b\tau}$.

The proof of the fact that $Y(s)=\cos(\lambda s)$ and $X(s)=\cos s$ are independent is trivial based on the theorems of M. Kac\cite{Kac1,Kac2,Kac3,Kac4}.

\section{Algebraic differential equations and chaotic solutions}
\label{Sec:AlDiffEqChao}

There exists a nontrivial fourth-\,order differential equations of type:
\begin{equation}
\label{EqGeny4}
P(y',y'',y''',y'''')=0,
\end{equation}
where $P$ is a polynomial in four variables, with integer coefficients, such that there exist solutions $y(t)$ with the property that $y(t)$ and $y(t+\tau)$ are statistically independent functions for most $\tau$.

A particular example of Eq.(\ref{EqGeny4}) is the following:
\begin{equation}
\label{EqSolExac1}
2y'y''-3y'y'''+y'y''''-y''y'''+{(y'')}^{2} =0.
\end{equation}

The function $y(t)=cos(e^{t})$ is a solution to this equation.

Here we should make a remark. The differential equations (\ref{EqExample}), (\ref{EqDuff1}), (\ref{EqDuff2}), (\ref{EqBriggs1}), and (\ref{EqSolExac1}) are similar in the sense that they are all of the type (\ref{EqGeneral}) or (\ref{EqGeny4}), and that all the terms are nonlinear.

We have shown that the equation (\ref{EqSolExac1}) possesses entire solutions as the following $y(t)=\cos(e^{t})$ which has the property that $y(t)$ and $y(t+\tau)$ are not correlated. This solution is real-analytic on $(-\infty,\infty)$.
This result shows that equations of type (\ref{EqGeny4}) can generate high complexity.

\section{Universal functions}
In this section we will follow the approach to universal equations developed by R. C. Buck \cite{BuckR} and M. Boshernitzan \cite{Bosher2}. In this paper, a family of functions will be called universal if these functions are dense in the space $C$[{\itshape \bfseries  I}] of all continuous real functions on the interval {\itshape \bfseries  I}$\subset\mathbb{R}$.
R. C. Buck has obtained universal partial algebraic differential equations using a very deep method: Kolmogorov's solution of Hilbert's Thirteen Problem. He has found that solutions to a smooth PDE can be dense in $C$[{\itshape \bfseries  I}]\cite{BuckR}.

The advantage of this approach is that we can construct universal functions which are real-\,analytic on $\mathbb{R}=(-\infty,\infty)$. Thus, the corresponding universal equations will have solutions that are real-\,analytic on $\mathbb{R}=(-\infty,\infty)$.

For any interval {\itshape \bfseries  I}$\subset \mathbb{R}=(-\infty,\infty)$, $C$[{\itshape \bfseries  I}] denotes the Banach space of real-\,valued continuous bounded functions on {\itshape \bfseries  I}.

Boshernitzan \cite{Bosher2} has studied the following families of functions:
\begin{equation}
\label{EqIntFam1}
y(t)=\int\limits_{0}^{t+a}\frac{bd}{1+d^{\,2}-\cos(bs)}\cos (e^{s})ds + c,
\end{equation}
where $d>0$, $a$, $b$, and $c$ are real parameters.

Another important family of functions is defined as follows:
\begin{equation}
\label{EqIntFam2}
y(t)=b\,n\int\limits_{0}^{t+a}\cos^{2n^{2}}(bs)\cos(e^{s})ds + c,
\end{equation}
where $a$, $b$, $c$ are parameters and $n \geq 1$ is an integer constant.

\begin{figure}[ht]
\centerline{
\includegraphics[height=5.2cm]{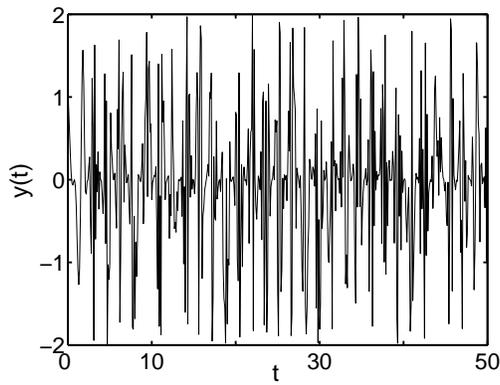}\hspace{0.1cm}}
\caption{\label{fig:FunFamBosher2} Time-\,series generated by (\ref{EqIntFam2}).}
\end{figure}
We should stress that all the functions in the family given by Eq.(\ref{EqIntFam2}) are real-analytic on $\mathbb{R}=(-\infty,\infty)$ and entire.

Boshernitzan has proved that each of the families of functions, (\ref{EqIntFam1}) and (\ref{EqIntFam2}), is dense in $C$[{\itshape \bfseries  I}], for any compact interval {\itshape \bfseries  I}$=[a,b]\subset\mathbb{R}$.

So these functions are universal. Hence it is possible to construct universal systems using these universal families of functions.

Other results of Boshernitzan are the following. There exists a nontrivial sixth-\,order algebraic differential equation of the form
\begin{equation}
P(y',y'',\cdots,y^{(6)})=0,
\end{equation}
such that any functions in the family (\ref{EqIntFam1}) is a solution. And there exists a nontrivial seventh-\,order algebraic differential equation
of the form
\begin{equation}
P(y',y'',\cdots,y^{(7)})=0,
\end{equation}
such that any function in the family (\ref{EqIntFam2}) is a solution.

From these theorems one can obtain an important results for us: There exists a nontrivial seventh-\,order differential equation, the real-\,analytic entire solutions of which are dense in $C$[{\itshape \bfseries  I}], for any compact interval {\itshape \bfseries  I}\;$\subset\mathbb{R}$. Thus this equation is universal. Boshernitzan has not constructed this equation explicitly.

The possibility of an entire approximation for any dynamics is very promising for practical applications.

\section{Universal systems of differential equations}

Using the inverse problem techniques and theorems of the works\cite{BuckR,Duffin,Rubel1,Rubel2,Bosher1,Bosher2,Elsner,Rubel3,Rubel4}, and the results contained in the present paper, it is possible to write down a system of differential equations such that any function of the family (\ref{EqIntFam2}) is a solution:
\begin{align}
\label{EqUniSyst1A}
P_{1}[x,x',\cdots,x''']&=0,\\
\label{EqUniSyst1B}
P_{2}[y,y',\cdots,y''']&=0,\\
\label{EqUniSyst1C}
z'=A\,x\,y.&
\end{align}

Eq. (\ref{EqUniSyst1A}) is constructed in such a way that all the functions $x=a[\cos(bx+c)]^{m}$ are solutions. The specific polynomial $P_{1}$ is
\begin{equation}
P_{1}=mx'''x^{2}-(3m-2)x''x'x + (2m-2){x'}^{3},
\end{equation}
where $m=2n^{2}>2$.

Eq.(\ref{EqUniSyst1B}) is constructed in such a way that $y(t)=\cos(e^{t})$ is a solution. The specific polynomial $P_{2}$ is defined as
\begin{equation}
P_{2}=yy'''-3yy''-y'y''+y'^{2}+2y'y.
\end{equation}
The vector solution $(x,y,z)$ to the system of equations (\ref{EqUniSyst1A}-\ref{EqUniSyst1C}) is such that for the variable $z(t)$ the functions (\ref{EqIntFam2}) are solutions.

Thus, the system of differential equations (\ref{EqUniSyst1A}-\ref{EqUniSyst1C})  can generate, in variable $z(t)$, universal functions.

Note that this is a seventh-\,order dynamical system as was predicted by Boshernitzan\cite{Bosher2}.

The system of equations (\ref{EqUniSyst1A}-\ref{EqUniSyst1C}) is presented in this paper for the first time.

\section{Noisy functions}
\label{Sec:NoisyFunc}
In this section we use several concepts and results from probability theory and mathematical statistics which can be consulted for instance in the books\cite{Book1, Book2}.

In sections (\ref{Sec:StatIndep}) and (\ref{Sec:AlDiffEqChao}) we discussed the function $y(t)=\cos(e^{t})$.
Note that the statistical independence between functions $y_{1}(t)=\cos(e^{t})$ and $y_{2}(t)=\cos(e^{t+\tau})$ implies the following relationship:

\begin{equation}
\label{EqExpecValueConj}
E [y_{1}^{k_{1}}(t)y_{2}^{k_{2}}(t)]= E [y_{1}^{k_{1}}(t)] E [y_{2}^{k_{2}}(t)],
\end{equation}
for all positive integers $k_{1}$ and $k_{2}$.

Here $E[x(t)]$ is the expected value of quantity $x(t)$. It can be calculated as follows:
\begin{equation}
\label{EqExpecValue}
E[x(t)]=\lim_{T\to\infty} \frac{1}{T}\int\limits_{0}^{T} x(t)dt.
\end{equation}

As $E[y_{1}(t)]=0$, we obtain that
\begin{equation}
E[y(t)y(t')]=0,
\end{equation}
for $t\neq t'$.

In fact, a direct calculation of the autocorrelation function confirms this results:
\begin{equation}
C(\tau)=\lim_{T\to\infty}\frac{1}{T}\int\limits_{0}^{T}\cos(e^{t})\cos(e^{t+\tau})dt=0
\end{equation}
for $\tau \neq 0$.

Another important statistical property of the independent functions $y_{1}(t)$ and $y_{2}(t)$ is
\begin{equation}
\eta(y_{1}(t),y_{2}(t))=\eta(y_{1}(t))\eta(y_{2}(t)),
\end{equation}
where $\eta(y)$ is the probability density.

In fact,
\begin{equation}
\label{Eq:ProbDensSquare1}
\eta(y_{1}) = \frac{1}{\pi\sqrt{1-y^{2}_{1}}},
\end{equation}
\begin{equation}
\label{Eq:ProbDensSquare2}
\eta(y_{2}) = \frac{1}{\pi\sqrt{1-y^{2}_{2}}},
\end{equation}
\begin{equation}
\label{Eq:ProbDensSquareJoint}
\eta (y_{1}, y_{2}) = \frac{1}{\pi^{2}\sqrt{(1-y^{2}_{1})(1-y^{2}_{2})}}.
\end{equation}

\begin{figure}
\centering
\subfloat[]{
\label{fig:EtaCos2D:a}
\includegraphics[width=0.45\linewidth]{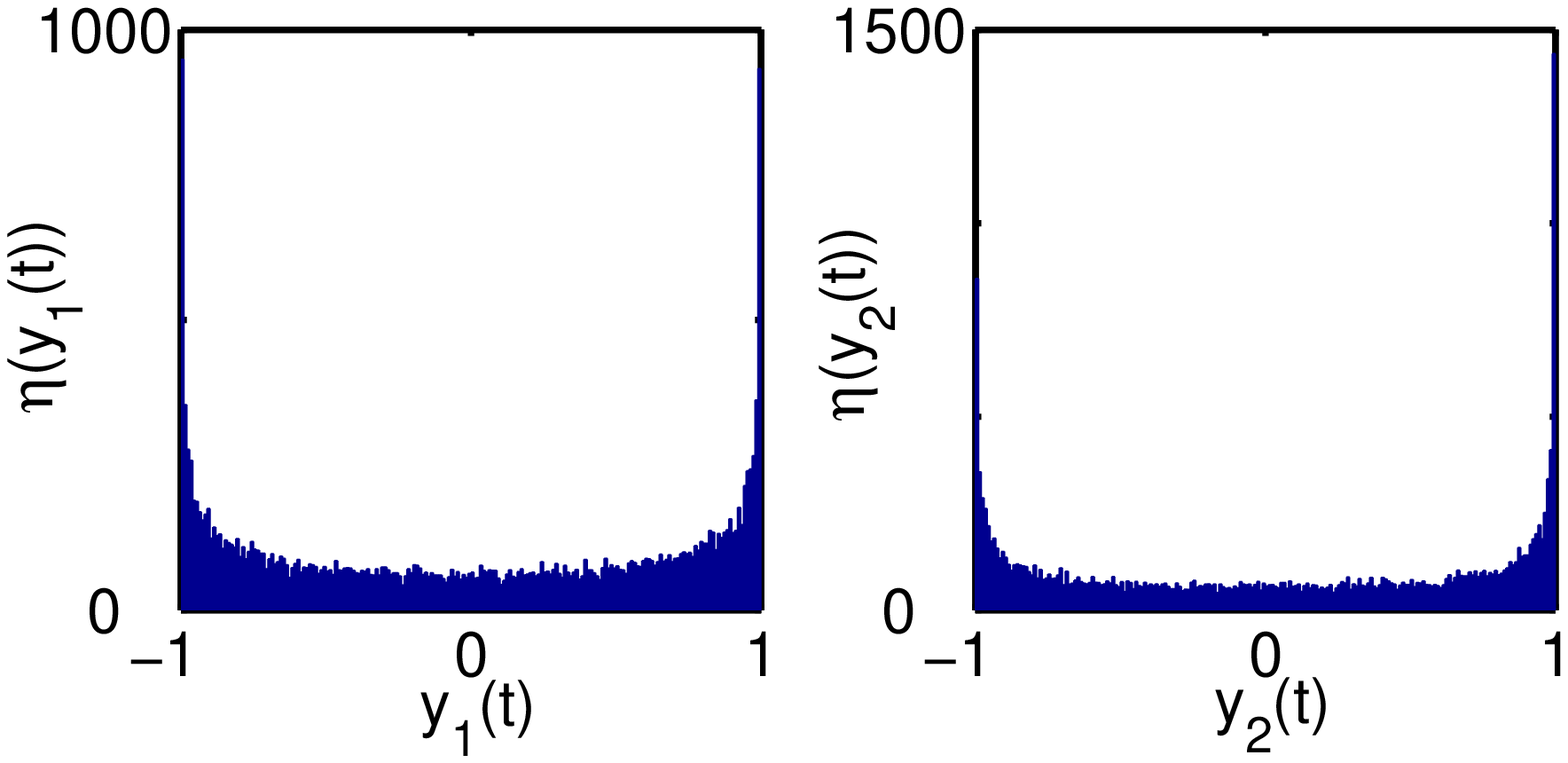}}
\\
\subfloat[]{
\label{fig:EtaCosD:b}
\includegraphics[width=0.35\linewidth]{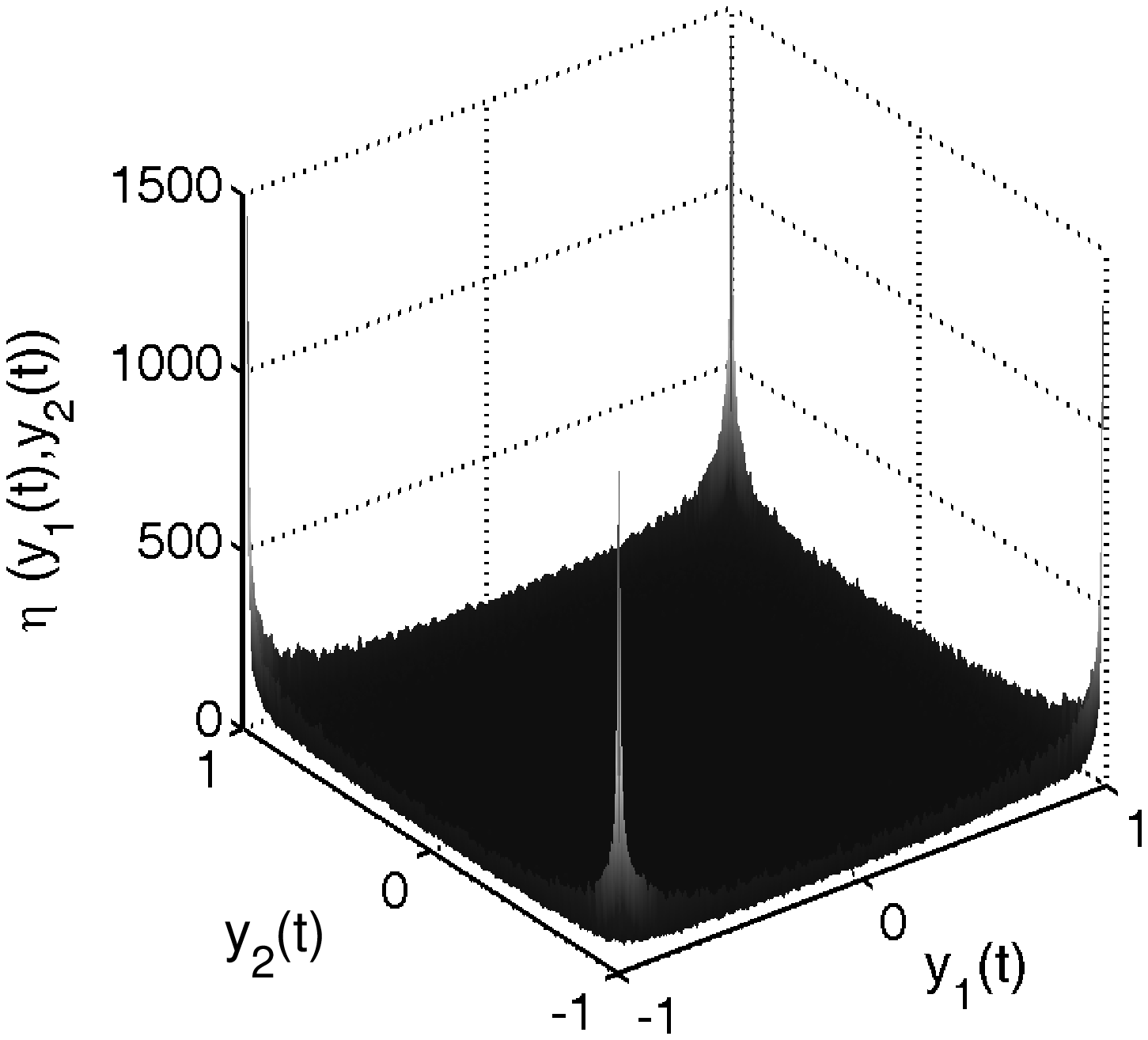}}
\quad
\caption{ \;(a)\; Probability density $\eta(y_{1})$ and $\eta(y_{2})$ ,\; (b)\; Probability density $\eta(y_{1},y_{2})$. }
\label{fig:EtaCosJoint}
\end{figure}

Fig.(\ref{fig:EtaCosJoint}) shows these properties.

Let us introduce a generalized function
\begin{equation}
\label{EqFunCont1}
y(t)=\cos(\phi(t)).
\end{equation}

We should remark here that the argument of function (\ref{EqFunCont1}), $\phi(t)$, does not need to be exponential all the time for $t\to\infty$,
in order to generate noisy dynamics.

In fact, it is sufficient for function $\phi(t)$ to be a bounded nonperiodic oscillating function which possesses repeating intervals with truncated exponential behavior\cite{GroupPhysD}.

A deep analysis of the Boshernitzan's proofs of the fact that the families functions (\ref{EqIntFam1}) and (\ref{EqIntFam2}) are dense in $C$[{\itshape \bfseries  I}] shows
that the random behavior of functions of type (\ref{EqFunCont1}) is crucial\cite{Bosher2}.

We are going to present here two  examples of these functions. The first is defined  by the equation
\begin{equation}
\label{Eq41}
x(t)= \cos\left\{A \exp \left[ a_{1}\sin(\omega_{1}t + \phi_{1})+ a_{2}\sin(\omega_{2}t + \phi_{2})+a_{3}\sin(\omega_{3}t+ \phi_{3})\right] \right\}.
\end{equation}

The second function is given by the equation
\begin{equation}
\label{Eq42}
\begin{split}
y(t)=&\cos \left\{ B_{1}\sinh\left[a_{1}\cos(\omega_{1}t + \phi_{1}) + a_{2}\cos(\omega_{2}t+ \phi_{2})\right]\right.\\& \left.+ B_{2}\cosh\left[a_{3}\cos(\omega_{3}t + \phi_{3})
+ a_{4}\cos(\omega_{4}t + \phi_{4})\right]\right\}.
\end{split}
\end{equation}

Note that if the frequencies $\omega_{i}$ are linearly independent over the rationals, in both cases, the argument function $\phi(t)$ is a nonperiodic function with truncated exponential behavior.

\begin{figure}[ht]
\centerline{
\includegraphics[height=5.2cm]{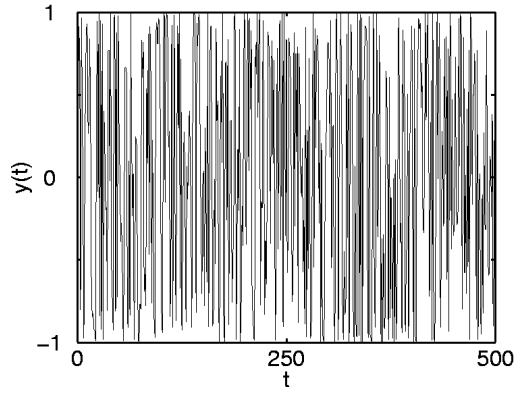}\hspace{0.1cm}}
\caption{\label{fig:FunCoshtA} Time-\,series generated by function (\ref{Eq42}).}
\end{figure}

Theoretical and numerical investigations give the result that, when the parameters satisfy certain conditions, they behave as the solutions to chaotic systems\cite{GroupChaos}. Figure (\ref{fig:FunCoshtA}) shows that the time-\,series generated by (\ref{Eq41}) is very complex.

Furthermore, they behave in such a way that $y_{1}=y(t)$ and $y_{2}=y(t+\tau)$ are statistically independent functions (in M. Kac's sense) for $\tau\neq0$. An illustration of these properties is that
\begin{equation}
\eta(y_{1},y_{2})=\eta(y_{1})\eta(y_{2}).
\end{equation}

Other noisy functions can be obtained using another generalization
\begin{equation}
x(t)=P[\phi(t)],
\end{equation}
where $P(y)$ is a general periodic function, and $\phi(t)$ is a nonperiodic exponential-like function as before.

The probability density of these functions depends on the choice of $P(y)$.

An important example is the following
\begin{equation}
\label{Eq45}
x(t)=\ln \left\{\tan^{2}[\phi(t)]\right\}.
\end{equation}

The probability density of the time-\,series produced by function (\ref{Eq45}) is a Gaussian-like law\cite{GroupPhysD}.

Another remarkable property of function (\ref{Eq45}) is the following:
\begin{equation}
E[x(t)x(t')]= D\delta(t-t'),
\end{equation}
where $E[x(t)x(t')]$ is defined as in Eq.(\ref{EqExpecValueConj}), and $\delta(t-t')$ is Dirac's delta-function.

Thus, function (\ref{Eq45}) possesses all the properties usually required in the stochastic equations with $\delta$-\,correlated noisy perturbations\cite{Hanggi,Gammaitoni}.

\section{Properties of chaotic solutions}
\label{SecChaoticSol}
A Lyapunov exponent of a dynamical system characterizes the rate of separation of infinitesimally close trajectories.

Suppose $\delta z_{0}$ is the initial distance between two trajectories, and $\delta z(t)$ is the distance between the trajectories at time $t$.

The maximal Lyapunov exponent can be defined as follows:
\begin{equation}
\lambda = \lim_{t\to\infty}\frac{1}{t}\ln\left|\frac{\delta z(t)}{\delta z_{0}}\right|.
\end{equation}

Consider the dynamical system
\begin{equation}
\frac{d\vec{x}}{dt}=\vec{F}(\vec{x}).
\end{equation}

The variational equation is
\begin{equation}
\frac{d\vec{\phi}}{dt}=D_{x}\vec{F}(\vec{x})\vec{\phi}(t).
\end{equation}

The Lyapunov exponent $\lambda$ satisfies the equation
\begin{equation}
\lambda = \lim_{t\to\infty}\frac{1}{t}\ln\left|\phi(t,x_{0})\right|.
\end{equation}

A solution will be considered chaotic if $\lambda>0$.

A related property of chaotic systems is sensitive dependence on initial conditions\cite{Guckenheimer,Jackson}.

Function
\begin{equation}
\label{EqSolE}
y(t)=\cos[\exp(t+\phi)]
\end{equation}
has been shown to be the solution to some dynamical system. Here $\phi$ can define the initial condition.

The Lyapunov exponent of solution (\ref{EqSolE}) can be calculated exactly $\lambda=\ln e=1>0$.

The functions (\ref{Eq41}), (\ref{Eq42}), and (\ref{Eq45}) are also chaotic solutions in this sense. The initial conditions are defined by the parameters $\phi_{1}$, $\phi_{2}$ and $\phi_{3}$.

Let us discuss sensitive dependence on initial conditions in the context of these functions.

Let $S$ be the set of functions defined by one of the families given by equations  (\ref{Eq41}), (\ref{Eq42}), (\ref{Eq45}), and (\ref{EqSolE}).

A set $S$ exhibits sensitive dependence if there is a $\delta$ such that for any $\epsilon>0$ and each $y_{1}(\phi,t)$ in $S$, there is a $y_{2}(\phi' ,t)$, also in $S$, such that $\left| y_{1}(\phi,0)-y_{2}(\phi',0)\right|<\epsilon$, and $\left| y_{1}(\phi,t)-y_{2}(\phi',t)\right|>\delta$ for some $t>0$.

The exponential behavior in the arguments of these functions ((\ref{Eq41}), (\ref{Eq42}), (\ref{Eq45}), and (\ref{EqSolE})) makes them chaotic (see a discussion in Ref. \cite{Suarez}).

All these functions possess equivalent dynamical and statistical properties.

Following Boshernitzan theory \cite{Bosher2} of modulo $1$ sequences, it is possible to construct the following families of universal functions:
\begin{equation}
\label{EqSolD}
y(t)=bn\int\limits_{0}^{t+a}\cos^{2n^{2}}(bs)x(s)ds + c,
\end{equation}
where $x(s)$ is one of the functions ((\ref{Eq41}), (\ref{Eq42}),(\ref{Eq45})).

\section{Dynamical systems with noisy solutions}

In this section we will construct autonomous dynamical systems, the solutions of which are the noisy functions discussed in the previous section.

Consider the following dynamical system
\begin{align}
\label{Eq47}
x_{1}'&=x_{1}[1-(x_{1}^{2}+y_{1}^{2})] -\omega_{1} y_{1},\\
\label{Eq48}
y_{1}'&=y_{1}[1-(x_{1}^{2}+y_{1}^{2})] +\omega_{1} x_{1},\\
\label{Eq49}
x_{2}'&=x_{2}[1-(x_{2}^{2}+y_{2}^{2})] -\omega_{2} y_{2},\\
\label{Eq50}
y_{2}'&=y_{2}[1-(x_{2}^{2}+y_{2}^{2})] +\omega_{2} x_{2},\\
\label{Eq51}
x_{3}'&=x_{3}[1-(x_{3}^{2}+y_{3}^{2})] -\omega_{3} y_{3},\\
\label{Eq52}
y_{3}'&=y_{3}[1-(x_{3}^{2}+y_{3}^{2})] +\omega_{3} x_{3},\\
\label{Eq53}
z'&=\left[a_{1}x_{1}+a_{2}x_{2}+a_{3}x_{3}\right]z,\\
\label{Eq54}
u'&=A\cos[\theta z].
\end{align}

Note that the equations (\ref{Eq47}-\ref{Eq52}) define three pairs of limit-cycle two-dimensional dynamical systems. The exact solutions to these limit-cycle systems are well-known.

If we define
\begin{equation}
Q(t)=a_{1}x_{1}(t)+a_{2}x_{2}(t)+a_{3}x_{3}(t),
\end{equation}
the function $Q(t)$ will be a quasiperiodic time-\,series.

Equation (\ref{Eq53}) will provide us with the appropriate nonperiodic truncated exponential behavior. Finally, the solution to equation (\ref{Eq54})
will have properties equivalent to these of function (\ref{Eq41}). The solution to Eq.(\ref{Eq54}) is chaotic in the sense discussed in section (\ref{SecChaoticSol}). The maximal Lyapunov exponent of this dynamical system is positive and the solutions possess sensitive dependence on initial conditions.

\begin{figure}
\centering
\subfloat[]{
\label{fig:UniEqAcopA1:a}
\includegraphics[width=0.35\linewidth]{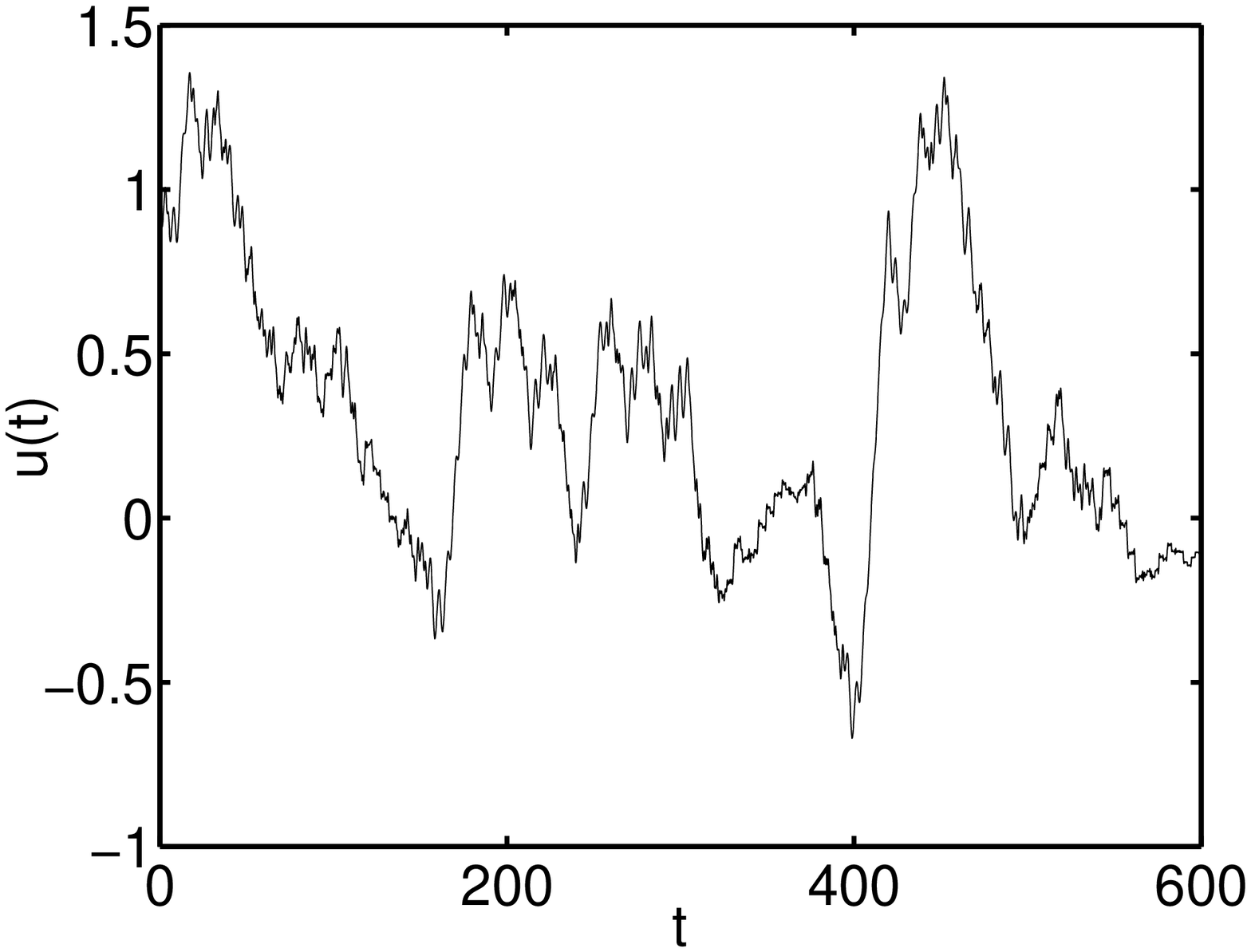}}
\quad
\subfloat[]{
\label{fig:UniEqAcopA3:b}
\includegraphics[width=0.34\linewidth]{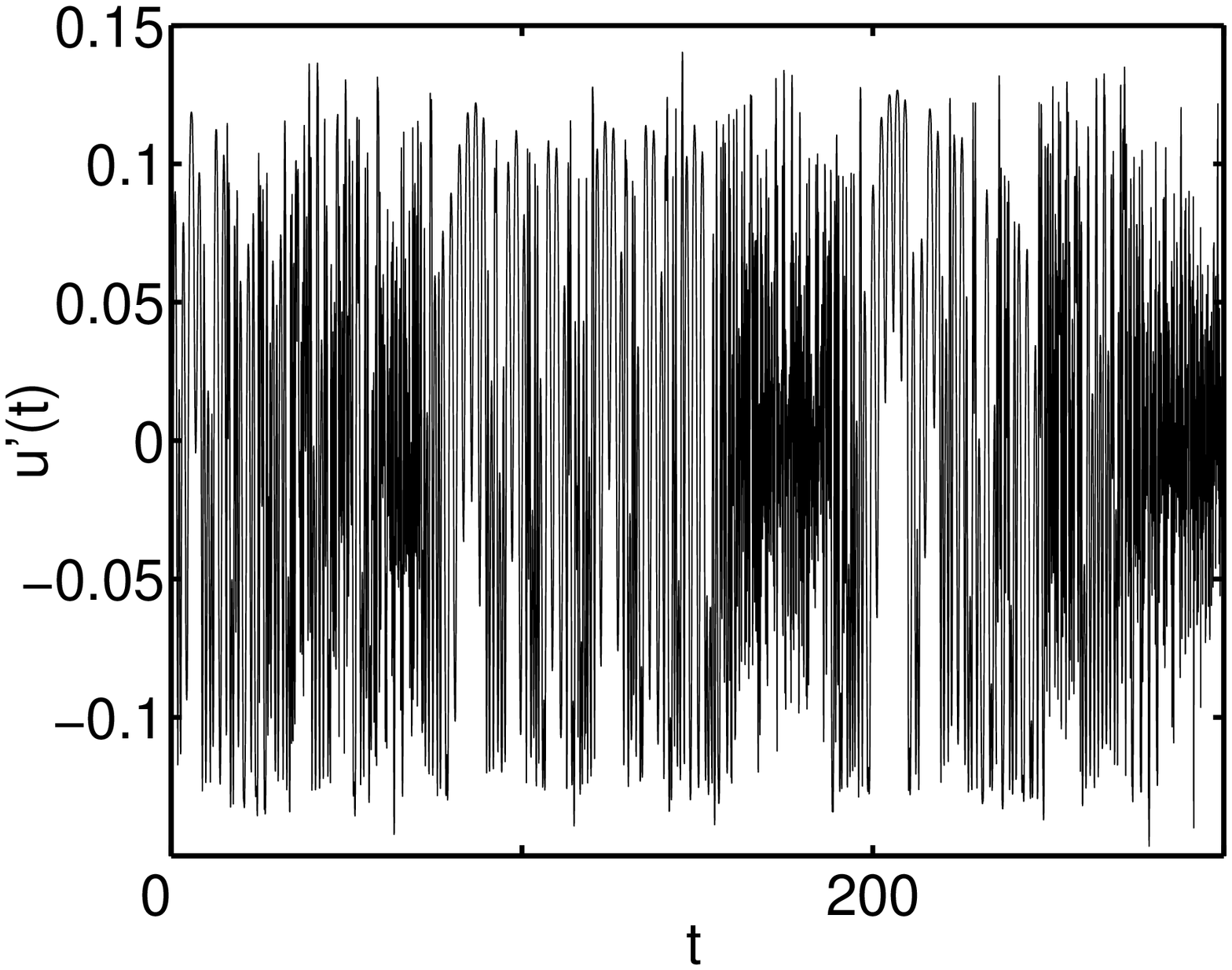}}
\\
\subfloat[]{
\label{fig:UniEqAcopDivA3:c}
\includegraphics[width=0.34\linewidth]{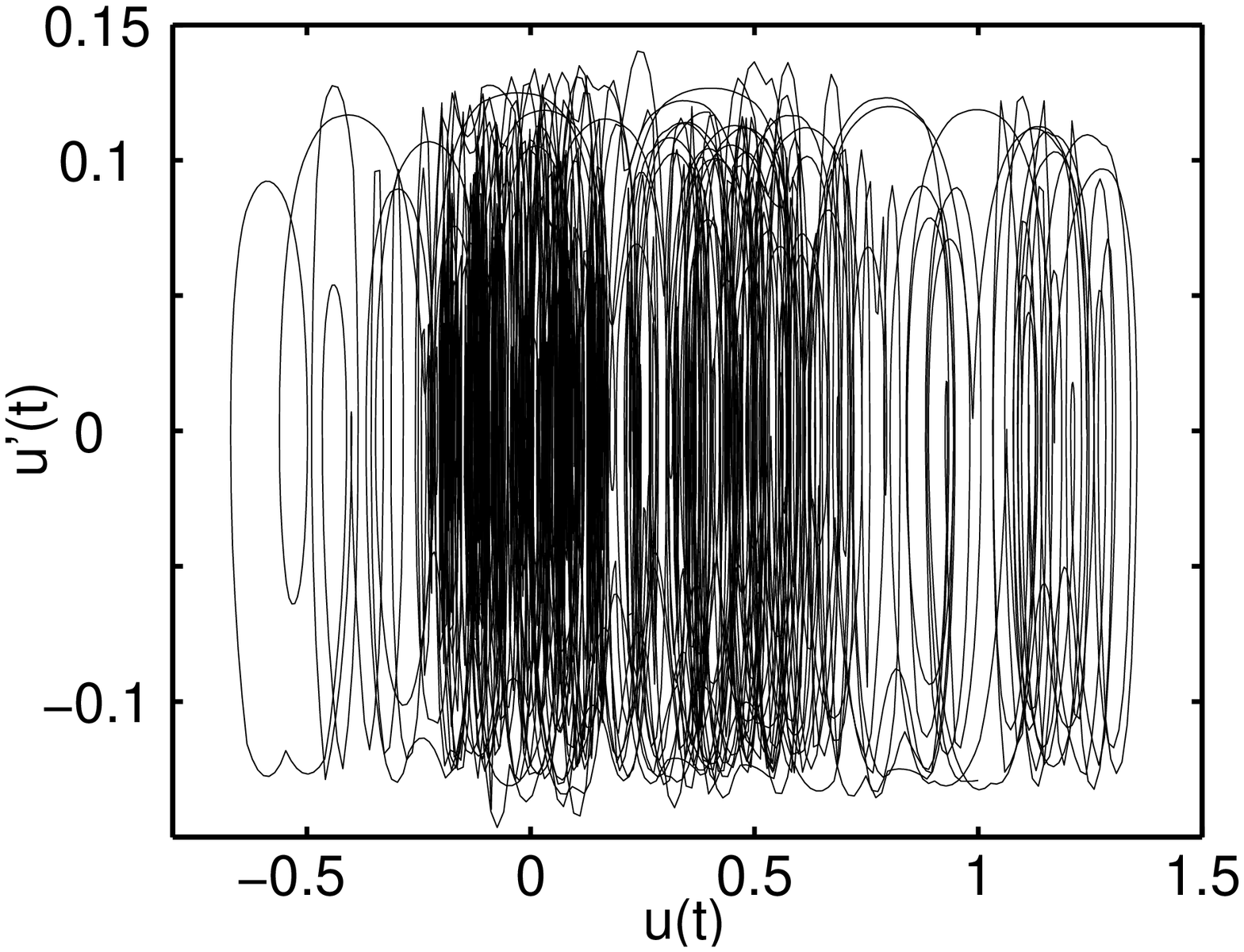}}
\caption{ \;(a)Time-\,series of variable $u(t)$,(b) $u'(t)$ and,\; (c)\; Phase portrait \; generated by  the autonomous dynamical system (\ref{Eq56}-\ref{Eq63}).}
\label{fig:FunAcopA1}
\end{figure}

If we need more variability in the solutions, we can construct a dynamical system such that the right-\,hand parts of the equations depend on function $u(t)$ which is known to be highly nonperiodic.

These ideas lead to the next autonomous dynamical system:
\begin{align}
\label{Eq56}
x_{1}'&=x_{1}[1-(x_{1}^{2}+y_{1}^{2})] -\omega_{1} y_{1}+\varepsilon_{1}u,\\
\label{Eq57}
y_{1}'&=y_{1}[1-(x_{1}^{2}+y_{1}^{2})] +\omega_{1} x_{1},\\
\label{Eq58}
x_{2}'&=x_{2}[1-(x_{2}^{2}+y_{2}^{2})] -\omega_{2} y_{2}+\varepsilon_{2}u,\\
\label{Eq59}
y_{2}'&=y_{2}[1-(x_{2}^{2}+y_{2}^{2})] +\omega_{2} x_{2},\\
\label{Eq60}
x_{3}'&=x_{3}[1-(x_{3}^{2}+y_{3}^{2})] -\omega_{3} y_{3}+\varepsilon_{3}u,\\
\label{Eq61}
y_{3}'&=y_{3}[1-(x_{3}^{2}+y_{3}^{2})] +\omega_{3} x_{3},\\
\label{Eq62}
z'&=\left[a_{1}x_{1}+a_{2}x_{2}+a_{3}x_{3}+a_{4}u\right]z,\\
\label{Eq63}
u'&=A_{1}\cos[\theta_{1} z]+A_{2}\cos[\theta_{2} z].
\end{align}
Now all the components of the solutions to system (\ref{Eq56}-\ref{Eq63}) are chaotic.

The solution to Eq.(\ref{Eq63}) is chaotic in the sense that the maximal Lyapunov exponent is positive (see section (\ref{SecChaoticSol})). In the dynamical system (\ref{Eq56}-\ref{Eq63}), the limit cycle subsystems (\ref{Eq56},\ref{Eq57}), (\ref{Eq58},\ref{Eq59}) and (\ref{Eq60},\ref{Eq61}) are now coupled to the chaotic component $u(t)$.

We now address the function (\ref{Eq42}). We have to construct a dynamical system with solutions that behave like this function.

Consider the autonomous dynamical system:
\begin{align}
\label{Eq64}
x_{1}'&=x_{1}[1-(x_{1}^{2}+y_{1}^{2})] -\omega_{1} y_{1},\\
\label{Eq65}
y_{1}'&=y_{1}[1-(x_{1}^{2}+y_{1}^{2})] +\omega_{1} x_{1},\\
\label{Eq66}
x_{2}'&=x_{2}[1-(x_{2}^{2}+y_{2}^{2})] -\omega_{2} y_{2},\\
\label{Eq67}
y_{2}'&=y_{2}[1-(x_{2}^{2}+y_{2}^{2})] +\omega_{2} x_{2},\\
\label{Eq68}
x_{3}'&=x_{3}[1-(x_{3}^{2}+y_{3}^{2})] -\omega_{3} y_{3},\\
\label{Eq69}
y_{3}'&=y_{3}[1-(x_{3}^{2}+y_{3}^{2})] +\omega_{3} x_{3},\\
\label{Eq70}
x_{4}'&=x_{4}[1-(x_{4}^{2}+y_{4}^{2})] -\omega_{4} y_{4},\\
\label{Eq71}
y_{4}'&=y_{4}[1-(x_{4}^{2}+y_{4}^{2})] +\omega_{4} x_{4},\\
\label{Eq72}
z_{1}'&=-\left(a_{1}\omega_{1}y_{1}+a_{2}\omega_{2}y_{2}\right)z_{2},\\
\label{Eq73}
z_{2}'&=-\left(a_{1}\omega_{1}y_{1}+a_{2}\omega_{2}y_{2}\right)z_{1},\\
\label{Eq74}
z_{3}'&=-\left(a_{3}\omega_{3}y_{3}+a_{4}\omega_{4}y_{4}\right)z_{4},\\
\label{Eq75}
z_{4}'&=-\left(a_{3}\omega_{3}y_{3}+a_{4}\omega_{4}y_{4}\right)z_{3},\\
\label{Eq76}
u'&=A\cos[B_{1}z_{1}+B_{4}z_{4}].
\end{align}

The explanation of this system is similar to that of equations (\ref{Eq47}-\ref{Eq54}).
\section{Universal functions and dynamical systems}

Based on the properties of functions (\ref{Eq21}), (\ref{EqIntFam1}), (\ref{EqIntFam2}), (\ref{Eq41}), and (\ref{Eq42}), we propose the following set of equations as a universal dynamical system:
\begin{align}
\label{Eq77}
x_{1}'&=x_{1}[1-(x_{1}^{2}+y_{1}^{2})] -\omega_{1} y_{1},\\
\label{Eq78}
y_{1}'&=y_{1}[1-(x_{1}^{2}+y_{1}^{2})] +\omega_{1} x_{1},\\
\label{Eq79}
x_{2}'&=x_{2}[1-(x_{2}^{2}+y_{2}^{2})] -\omega_{2} y_{2},\\
\label{Eq80}
y_{2}'&=y_{2}[1-(x_{2}^{2}+y_{2}^{2})] +\omega_{2} x_{2},\\
\label{Eq81}
x_{3}'&=x_{3}[1-(x_{3}^{2}+y_{3}^{2})] -\omega_{3} y_{3},\\
\label{Eq82}
y_{3}'&=y_{3}[1-(x_{3}^{2}+y_{3}^{2})] +\omega_{3} x_{3},\\
\label{Eq83}
x_{4}'&=x_{4}[1-(x_{4}^{2}+y_{4}^{2})] -\omega_{4} y_{4},\\
\label{Eq84}
y_{4}'&=y_{4}[1-(x_{4}^{2}+y_{4}^{2})] +\omega_{4} x_{4},\\
\label{Eq85}
z_{1}'&=-\left(a_{1}\omega_{1}y_{1}+a_{2}\omega_{2}y_{2}\right)z_{2},\\
\label{Eq86}
z_{2}'&=-\left(a_{1}\omega_{1}y_{1}+a_{2}\omega_{2}y_{2}\right)z_{1},\\
\label{Eq87}
z_{3}'&=-\left(a_{3}\omega_{3}y_{3}+a_{4}\omega_{4}y_{4}\right)z_{4},\\
\label{Eq88}
z_{4}'&=-\left(a_{3}\omega_{3}y_{3}+a_{4}\omega_{4}y_{4}\right)z_{3},\\
\label{Eq89}
u'&=\omega_{1}nx_{1}^{2n^{2}}\cos\left[B_{1}z_{1}+B_{4}z_{4}+a\right].
\end{align}

The solution to Eq.(\ref{Eq89}) is a function with all the properties of Boshernitzan's family of functions (\ref{EqIntFam2}). Thus this family of functions is dense in $C$[{\itshape \bfseries  I}].

The dynamical system (\ref{Eq77}-\ref{Eq89}) has been constructed  using the same technique developed in the papers \cite{BuckR,Bosher2} about universal differential equations. That is, a family of functions known to be dense in $C$[{\itshape \bfseries  I}] is utilized as the starting point for an inverse-problem procedure that consists in reconstructing differential equations the solutions of which are the functions that belong to the universal family of functions.

This dynamical system can be realized in practice using nonlinear circuits as discussed in Ref.\cite{GroupPhysLettA}.

\section{Conclusions}
We have discussed the concept of ``Universal functions" and their relevance to the theory of universal differential equations.

We believe that the method of construction of universal differential equations using universal functions is more powerful that the method based on splines.

We have found a connection between  the universal families of functions proposed in a very important paper by Boshernitzan \cite{Bosher2} and recently obtained  exact solutions  to chaotic systems.

We have shown that some functions $x(t)$ that are exact solutions to chaotic systems possess  the property that $y_{1}=x(t)$ and $y_{2}=x(t+\tau)$ are statistically independent functions in the sense of M. Kac.

We have constructed algebraic differential equations  that possess solutions with these properties. These equations have, in fact, solutions that behave like noise.

Some known universal equations can only approximate these functions using ``solutions" constructed with polynomial or trigonometric splines. The actual exact solutions to the differential equations are not ``noisy".

We have constructed a system of differential equations, the solutions of which are Boshernitzan's functions. Boshernitzan's functions are real analytic on $\mathbf{\mathbb{R}}=(-\infty,\infty)$. One of the families of Boshernitzan's functions are real-\,analytic  entire functions on $\mathbf{\mathbb{R}}=(-\infty,\infty)$.

We have developed universal-\,like functions that behave as $\delta$-\,correlated noise.

We have constructed physically realizable dynamical systems that possess solutions that are universal-like functions.

The theory of universal differential equations has been linked from the beginning with applications in analog computing\cite{Rubel1,Duffin,Rubel2,Bosher1,Bosher2,Rubel3,Rubel4}.

We believe that the present results can be of interest in the construction of real analog computers because the discussed  dynamical systems can be realized in practice using nonlinear circuits\cite{GroupPhysLettA,GroupPhysD,Suarez}.

They can also find applications in chaos-\,based secure communications technologies \cite{GroupPhysLettA,Group1}.

\end{document}